\UseRawInputEncoding
\documentclass[superscriptaddress, notitlepage, reprint]{revtex4-1}
\usepackage[english]{babel}
\usepackage{amsmath,amsthm}
\usepackage{amsfonts}
\usepackage[pdfborder={0 0 0}, colorlinks=true, urlcolor=blue, linkcolor=blue, citecolor=blue]{hyperref}
\usepackage{color}
\usepackage{graphicx}
\usepackage{float}
\usepackage{subfigure}
\usepackage{lipsum}
\usepackage{epstopdf}
\usepackage{dcolumn}
\begin{document}

\title{Thermometry with a Dissipative Heavy Impurity }
\author{Dong  Xie}
\email{xiedong@mail.ustc.edu.cn}
\affiliation{College of Science, Guilin University of Aerospace Technology, Guilin, Guangxi 541004, People's Republic of China}
\author{Chunling Xu}
\affiliation{College of Science, Guilin University of Aerospace Technology, Guilin, Guangxi 541004, People's Republic of China}

\begin{abstract}
Improving the measurement precision of low temperature is significant in fundamental science and advanced quantum technology application.
However, the measurement precision of temperature $T$ usually diverges as $T$ tends to 0. Here, by utilizing a heavy impurity to measure the temperature
of a Bose gas, we obtain the Landau bound to precision $\delta^2 T\propto T^2$ to avoid the divergence. Moreover, when the initial momentum
of the heavy impurity is fixed and non-zero, the measurement precision  can be $\delta^2 T\propto T^3$ to break the Landau bound. We derive the momentum distribution of the heavy impurity at any moment and obtain the optimal measurement precision of the temperature by calculating the Fisher information. As a result, we find that enhancing the expectation value of the initial momentum can help to improve the measurement precision. In addition, the momentum measurement is the optimal measurement of the temperature in the case of that the initial momentum is fixed and not equal to 0. The kinetic energy measurement is the optimal measurement in the case of that the expectation value of the initial momentum is 0.
Finally, we obtain that the temperatures of two Bose gases can be measured simultaneously. The simultaneous measurement precision is proportional to $T^2$ when two temperatures are close to $T$.

\end{abstract}
\maketitle

\section{Introduction}
With the rapid development of quantum technology, precise measurement of low temperature is becoming an important and significant subject in the field of quantum metrology\cite{lab1,lab2,lab3,lab4,lab5,lab6,lab7} and  quantum thermodynamics\cite{lab8,lab9,lab10}.  The measurement of low
temperature has always been a challenging task due to that the uncertainty of temperature diverges as the temperature tends to zero\cite{lab11,lab12}.
A strong coupling\cite{lab13}, periodic driving\cite{lab14} and correlations among multiple probes induced by the common
bath\cite{lab15} have been used to  slow down the divergence. In order to truly avoid the divergence, quantum system with a non-vanishing gap\cite{lab16}, a non-Markovian reservoir\cite{lab17}, and invariant subspaces due
to the polariton thermalization\cite{lab18} have been utilized to obtain the Landau bound to precision $\delta^2 T\propto T^2$. In these cases, the measurement uncertainty is not only not divergent but also getting smaller and smaller as the temperature decreases to 0.

With the progress of experimental techniques in quantum gases, the high-resolution imaging of
distinguishable impurities has been realized\cite{lab19,lab20,lab21}. The strength of
the impurity-environment coupling can be tuned to obtain many significant physical results.
The motion of mobile impurities through one-dimensional quantum liquid or gas has been extensively studied\cite{lab22,lab23,lab24,lab25,lab26}.
A heavy impurity moving through a Luttinger liquid\cite{lab27} was explored to find that
the friction force experienced by the impurity behaves as the
fourth power of temperature ($T^4$). By controlling the system parameter, the friction force can dramatically
change its temperature dependence from $T^4$ to $T^8$ in Bose liquid\cite{lab28,lab29}. In recent work\cite{lab30}, the low-temperature $T^2$
dependence of the friction force has been obtained for a strong coupling between a heavy impurity and a Bose gas, which is contrasted with
the expected $T^4$ scaling for a weak coupling. In order to obtain the friction force from the Bose gas, the temperature of the Bose gas should be measured accurately in advance.

In this article, we use a heavy impurity to measure the temperature of a Bose gas. Based on that the relation between the friction force and the momentum is linear,  we analytically derive the momentum distribution of the heavy impurity at any time given by the initial Gaussian distribution. The optimal temperature measurement precision can be analytically derived by achieving the Fisher information from the momentum distribution of the heavy impurity. As a result, we show that
the Landau bound to precision $\delta^2 T\propto T^2$ can be obtained in the general case.  More importantly, when the initial momentum
of the heavy impurity is fixed and non-zero, the measurement precision  can be $\delta^2 T\propto T^3$ to break the Landau bound. And we find that enhancing the expectation value of the initial momentum can help to improve the measurement precision.
In the case of that the initial momentum is fixed and not equal to 0, we show that the momentum measurement is the optimal measurement of the low temperature. When the initial momentum is not fixed and the expectation value is not equal to 0, the measurement precision proportional to $T^2$ can still be obtained by using the momentum measurement. The the kinetic energy measurement just happens to be the optimal measurement in the case of that the expectation value of the initial momentum is 0.
Finally, we  show that the simultaneous measurement precision is proportional to $T^2$ when the temperatures of two gases are close to $T$.

This article is organized as follows. In Section II, we introduce the model of the heavy impurity in the Bose gas. In Section III, the momentum distribution of the heavy impurity is derived given by the initial Gaussian distribution. In Section IV, the optimal estimation precision is obtained by the Fisher information. In Section V, two practical measurements are used to obtain the temperature measurement precision. The simultaneous estimation of two temperatures is discussed in Section VI. We make a conclusion and a discussion on the feasibility of the experiment in Section VII.

 \section{Heavy Impurity in a Bose Gas}
We consider the system of a mobile heavy impurity in a Bose gas composed of one-dimensional interacting bosons. The total Hamiltonian is described by\cite{lab31}
\begin{align}
H_{T}=H_B-\int dx\hat{\Psi}^\dagger\frac{\hbar^2}{2M}\partial_x^2\hat{\Psi}+G\int dx\hat{\Psi}^\dagger\hat{\Psi}\hat{\psi}^\dagger\hat{\psi},
\label{eq:1}
\end{align}
where $H_B$ denotes the Lieb-Liniger Hamiltonian of the Bose gas
\begin{align}
H_B=-\int dx\hat{\psi}^\dagger\frac{\hbar^2}{2m}\partial_x^2\hat{\psi}+g\int dx\hat{\psi}^\dagger\hat{\psi}^\dagger\hat{\psi}\hat{\psi},
\label{eq:2}
\end{align}
$\hat{\Phi}(x)=\Psi(x)\ [\psi(x)]$ denotes the bosonic field operator for the heavy impurity with the mass $M$ [ the single boson with the mass $m$], which satisfies the commutation relations $[\hat{\Phi}^\dagger(x),\hat{\Phi}^\dagger(x')]=\delta(x-x')$ and $[\hat{\Phi}(x),\hat{\Phi}(x')]=0$ with $\delta(x-x')$ being the Dirac delta function. In  the Bose gas,  the contact interaction between bosons are repulsive and weak, i.e., $\hbar^2n_0/m\gg g>0$ with $n_0$ being the mean density of the bosons. In the case of the weak interaction between bosons, the quasi-particles of the Hamiltonian $H_B$ have the Bogoliubov dispersion relation\cite{lab32} $\epsilon_P=\sqrt{v^2p^2+p^4/4m^2}$ with the sound velocity $v=\sqrt{gn_0/m}$.
And $G$ denotes the density-density interaction strength between the heavy impurity and the Bose gas.

\section{Distribution function of the impurity}
Due to that the heavy impurity collides with thermally excited bosons in the Bose gas, the heavy impurity motion is stochastic. The momentum distribution function $f(t,P)$
of the heavy impurity ($M\gg m$) can be characterized by the Fokker-Planck form\cite{lab33,lab34}
\begin{align}
\frac{\partial f(t,P)}{\partial t}=\frac{\partial }{\partial P}[-F f(t,P)+\frac{1}{2}\frac{\partial Df(t,P) }{\partial P}],
\label{eq:3}
\end{align}
where $P$ represents the momentum of the heavy impurity. $F$ denotes the friction force due to scattering off thermally excited quasiparticles in the Bose gas, which is
given by
\begin{align}
F=-\frac{m^2v^2P}{2\pi\hbar M\tilde{T}}\int_0^\infty dk \frac{k^2|r(k,\tilde{G})|^2(2+k^2)}{\sinh^2(k\frac{\sqrt{4+k^2}}{4\tilde{T}})\sqrt{4+k^2}},
\label{eq:4}
\end{align}
where $\tilde{T}=\frac{T}{m v^2}$, and $\tilde{G}=\frac{G}{\hbar v}$. $r(k,\tilde{G})$ denotes the reflection amplitude of the Bogoliubov quasiparticles scattering off a heavy impurity, which has been studied by the Bogoliubov-de Gennes theory\cite{lab34a}. At low temperatures, $\tilde{T}\ll 1$, in the cases of weak coupling ($\tilde{G}\ll1$) and strong coupling $\tilde{G}\gg1$  the friction force $F$ can be simplified as\cite{lab30}
\begin{align}
F=-\Gamma({G})PT^n.
\label{eq:5}
\end{align}
For $1/\tilde{G}\ll\tilde{T}\ll1$, $n=2$ and $\Gamma({G})=\frac{2\pi}{3\hbar Mv^2}$. For $\tilde{T}\ll1/\tilde{G}\ll1$, $n=4$ and $\Gamma({G})=\frac{8\pi^3G^2}{15\hbar^3m^2 Mv^8}$. For $\tilde{G}\ll1\ \textmd{and} \ \tilde{T}\ll1$, $n=4$ and $\Gamma({G})=\frac{2\pi^3G^2}{15\hbar^3m^2 Mv^8}$.
And $D$ is the impurity diffusion coefficient, which is given by
\begin{align}
D=-2FMT/P.
\label{eq:6}
\end{align}

When the friction force $F$ is proportional to the momentum $P$ as shown in Eq.~(\ref{eq:5}), the Fokker-Planck form  in Eq.~(\ref{eq:3}) can be solved analytically. In general, the initial momentum distribution $f(0,P)$ is Gaussian, which is described by
\begin{align}
f(0,P)=\frac{1}{\sqrt{\pi\Delta}}\exp[-\frac{(P-P_0)^2}{\Delta}],
\label{eq:7}
\end{align}
where $\Delta/2$, $P_0$ represent the variance and the expectation value of the initial momentum, respectively.

By analytically solving the Fokker-Planck equation in Eq.~(\ref{eq:3}), we can obtain the momentum distribution at time $t$ with the initial Gaussian momentum distribution as shown in Eq.~(\ref{eq:7})
 \begin{align}
f(t,P)&=\frac{1}{\sqrt{2MT(1-e^{-2t/\tau}(1-\Delta'))}}\nonumber\\
&\times\exp[-\frac{(P-P_0e^{-t/\tau})^2}{2MT(1-e^{-2t/\tau}(1-\Delta'))}],
\label{eq:8}
\end{align}
where $\Delta'=\Delta/(2MT)$ and $\tau=1/(\Gamma T^n)$ with the abbreviation $\Gamma\equiv\Gamma(G)$ throughout the rest of this article.
\section{Fisher Information from Gaussian Momentum Distribution}
According to the Cram\'{e}r-Rao bound\cite{lab35}, the measurement precision of the temperature can be given by
\begin{align}
(\delta T)^2\geq\frac{1}{N\mathcal{F}[T]},
\label{eq:9}
\end{align}
where $N$ represents the total number of repeated experiments. Since the content of our next study has nothing to do with the number of measurements, we simply set $N=1$. $\mathcal{F}[T]$ denotes the Fisher information of the temperature $T$, which is described by
\begin{align}
\mathcal{F}(T)=\int_{-\infty}^\infty dP\frac{[\partial_T f(t,P)]^2}{f(t,P)},
\label{eq:10}
\end{align}
where the partial derivative $\partial_T f(t,P)=\frac{\partial f(t,P)}{\partial T}$.

When $\Delta$ is not an infinitesimal quantity, i.e., $\Delta>0$, and $t$ is not infinite, we can obtain the Fisher information in the low temperature limit $T^nt\Gamma\ll1$
\begin{align}
\mathcal{F}(T,\Delta>0)=\frac{2(nt\Gamma)^2 T^{2n-2}(P_0^2+\Delta)}{\Delta}.
\label{eq:11}
\end{align}
From the above equation, we can see that the Fisher information can be increased by increasing the initial expectation values of momentum.
When $\tilde{T}\ll1/\tilde{G}$, the coefficient $\Gamma$ increases by 4 times with the coupling strength,i.e., $\Gamma(\tilde{G}\gg1)=4\Gamma(\tilde{G}\ll1)$. Namely, the Fisher information increase 4 times with the coupling strength. When the coupling strength continues to increase until $\tilde{G}\gg1/\tilde{T}$,
the power of temperature goes from $n=4$ to $n=2$. Hence, in the case of $\tilde{G}\gg1/\tilde{T}$, the Fisher information is proportional to the square of the temperature ($\mathcal{F}(T)\propto T^2$). And it is independent of the coupling strength, which shows that the measurement precision of temperature will not be improved when the coupling strength increases to a certain extent. As the temperature $T$ approaches to 0, the Fisher information $\mathcal{F}(T)$ will approach to 0. It means that the uncertainty of the temperature $T$ is divergent at $T=0$.

When $\Delta=0$ and $t$ is not infinite, the Fisher information in the low temperature limit $T^nt\Gamma\ll1$ is given by
\begin{align}
\mathcal{F}(T,\Delta=0)=\frac{(1+n)^2}{2T^2}+\frac{n^2P_0^2T^{n-3}t}{2M}.
\label{eq:12}
\end{align}
When $T$ is close to 0, the Fisher information, $\mathcal{F}(T)\approx \frac{25}{2T^2}$, will be close to infinite.
This result is significant, which shows that the divergent question of the uncertainty at $T=0$ is completely solved by using a defined initial momentum, i.e., $\Delta=0$.
In the case of a fixed initial momentum, $f(0,P)=\delta(P-P_0)$, the uncertainty  of $T$ will be close to 0 as $T$ is close to 0. A simple summary is that reducing the uncertainty of the initial momentum distribution can effectively improve the measurement precision of the temperature, especially when the uncertainty of the initial momentum distribution is 0, the measurement precision of the temperature changes from infinity to 0 at the temperature $T=0$.

Next, we consider that the measurement time $t$ can be arbitrarily large. By optimizing the measurement time to obtain a better measurement precision, we consider that the measurement time is the characteristic time of the distribution function $f(t,P)$, i.e., $t=\tau=1/(\Gamma T^m)$. When $T$ is close to 0 and $\Delta>0$, the Fisher information is given by
\begin{align}
\mathcal{F}_{\tau}(\Delta>0)=\frac{2n^2(P_0^2+\Delta)}{T^2\Delta}.
\label{eq:13}
\end{align}
Contrary to the result in Eq.~(\ref{eq:5}), we find that $\mathcal{F}_{\tau}(\Delta>0)$ becomes larger and larger as the temperature $T$ decreases.
The essential reason is that the measurement precision of temperature close to 0 is improved by using the time resource tending to be infinite.

When $T$ is close to 0, $\Delta=0$ and $P_0\neq0$, the Fisher information is given by
\begin{align}
\mathcal{F}_{\tau}(\Delta=0)=\frac{n^2P_0^2}{MT^3(e^2-1)}.
\label{eq:14}
\end{align}
\begin{figure}[h]
\includegraphics[scale=0.9]{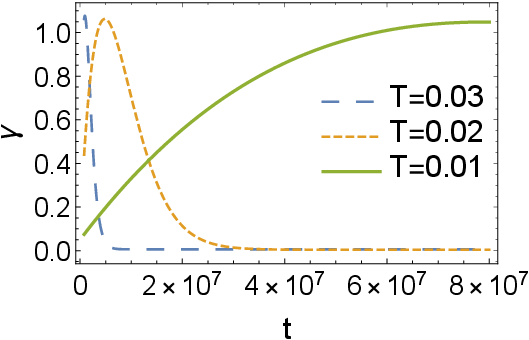}
 \caption{\label{fig.1}Evolution diagram of the ratio of the Fisher information at different low temperatures. Here, $\gamma$ denotes the ratio of the Fisher information $\mathcal{F}(T)$ at time $t$ and the Fisher information $\mathcal{F}_{\tau}(\Delta=0)$ at the characteristic time $\tau$. The dimensionless parameters chosen are given by: $P_0=1$, $M=1$, $n=4$, and $\Gamma=1$.}
\end{figure}
As shown in Fig.~\ref{fig.1}, we can see that the maximum ratio $\gamma=\mathcal{F}(T)/\mathcal{F}_{\tau}(\Delta=0)$ approaches to 1. It means that $\mathcal{F}_{\tau}(\Delta=0)$ is close to the optimal Fisher information. When $\Delta=0$, the characteristic time $\tau$ is close to the optimal measurement time at low temperature.
Comparing  Eq.~(\ref{eq:12})  and Eq.~(\ref{eq:14}), the Fisher information changes from being proportional to $1/T^2$ to $1/T^3$ in the case of $P_0\neq0$.
This is also due to the use of time resources that tend to be infinite.

When the initial momentum $P_0$ is 0, the Fisher information at the characteristic time $\tau$ is given by
 \begin{align}
\mathcal{F}_{\tau}(\Delta=0,P_0=0)=\frac{(2n+e^2-1)^2}{2T^2(e^2-1)^2}.
\label{eq:15}
\end{align}
This result shows that the initial non-zero momentum ($P_0$) can help to effectively use the time resources, and thus improve the measurement precision of the temperature.

\section{temperature measurement with momentum}
In the previous section, we give the optimal measurement precision for a given probability distribution by the Fisher information. Then we will use the error propagation formula to show whether the specific measurement can approach the optimal measurement result.

Firstly, we obtain the information of the temperature $T$ by the measurement of the momentum $P$. Given by the momentum distribution in Eq.~(\ref{eq:2}), we achieve the expectation value $\bar{P}$ and the variance $\delta^2P={\bar{P^2}}-{\bar{P}}^2$, which are described by
 \begin{align}
{\bar{P}}&=P_0e^{-t/\tau},\\
\delta^2P&=MT[1-e^{-2t/\tau}(1-\Delta')].
\end{align}
The uncertainty of the temperature $T$  can be derived by the error propagation formula\cite{lab36}

 \begin{align}
\delta^2T=\frac{\delta^2P}{|\partial_T{\bar{P}}|^2}
=\frac{MT(1-e^{-2t/\tau}(1-\Delta'))}{(P_0nt\Gamma)^2T^{2n-2}e^{-2t/\tau}}.
\end{align}

From the above equation, we can derive that the optimal measurement time can be close to the characteristic time $\tau$. At time $\tau$, the corresponding measurement precision of the temperature $T$ obtained by the measurement of momentum can be given by
 \begin{align}
\delta^2T|_{t=\tau}=\frac{MT(1-e^{-2}(1-\Delta'))}{(P_0n)^2T^{-2}e^{-2}}\\
\approx\frac{(e^2-1)MT^3+ T^2\Delta}{(P_0n)^2}
\end{align}

When $P_0\neq0$ and $\Delta=0$, we can find that $\delta^2T|_{t=\tau}=1/\mathcal{F}_{\tau}(\Delta=0)$. It shows that the momentum measurement is the optimal measurement of the low temperature in the case of that the initial momentum is fixed ($\Delta=0$) and not equal to 0 ($P_0\neq0$). When the initial momentum is not fixed ($\Delta>0$) and the expectation value is not equal to 0 ($P_0\neq0$), the measurement precision proportional to $T^2$ can still be obtained by using the momentum measurement. This means that the momentum measurement is close to the optimal measurement as $T$ tends to 0 in the
case of $P_0\neq0$ and $\Delta>0$.

However, when $P_0=0$, one can obtain that $\delta^2T|_{t=\tau}\rightarrow\infty$. It shows that the measurement of momentum is the worst measurement, which can not obtain any information of the temperature $T$ in the case of $P_0=0$.
\subsection{Temperature measurement with kinetic energy}
 In order to deal with the case of $P_0=0$, we try to use the kinetic energy measurement
$P^2/(2M)$. The expectation value and the variance of $P^2$ are given by
 \begin{align}
{\bar{P^2}}&=MT[1-e^{-2t/\tau}(1-\Delta')];\\
\delta^2P^2&=2M^2T^2[1-e^{-2t/\tau}(1-\Delta')]^2.
\end{align}

By substituting the above equations into the error propagation formula, the measurement precision of the temperature is derived

\begin{align}
\delta^2T=\frac{T^2[2(e^{2t/\tau}-1)MT+\Delta]^2}{2[MT(e^{2t/\tau}-1)+nt(2MT-\Delta)/\tau]^2}.
\end{align}
For $t\rightarrow\infty$, the measurement precision of the low temperature obtained by the kinetic energy measurement is $\delta^2T=2T^2$, which is consistent with the results obtained by
the Fisher information in the previous section.
For $t=\tau$ and $\Delta=0$, the measurement precision of the low temperature is $\delta^2T=\frac{2(e^2-1)^2T^2}{(e^2-1+2n)^2}$, which is consistent with the result as shown in Eq.~(\ref{eq:15}).
For $\Delta>0$, the measurement precision of the low temperature is $\delta^2T=\frac{T^2}{2n^2}$, which is also consistent with the result  by
the Fisher information in Eq.~(\ref{eq:13}).
Therefore, in the case of $P_0=0$, the temperature measurement with the kinetic energy is the optimal measurement.

\section{simultaneous measurement of two temperatures}
We consider that the heavy impurity passes through the first Bose gas with temperature $T_1$ and the second Bose gas with temperature $T_2$.
Given the initial Gaussian momentum distribution of the Heavy impurity as shown in Eq.~(\ref{eq:7}), the momentum distribution at time $t_1+t_2$ is described by
 \begin{align}
f(t_1+t_2,P)=&\frac{1}{\sqrt{2MT_2[1-e^{-2t_2/\tau_2}(1-\frac{\Delta_1}{2MT_2})]}}\times\nonumber\\
&\exp[-\frac{(P-P_0e^{-t_1/\tau_1-t_2/\tau_2})^2}{2MT_2[1-e^{-2t_2/\tau_2}(1-\frac{\Delta_1}{2MT_2})]}],
\end{align}
where $t_1$ ($t_2$) is the interaction time between the heavy impurity and the first (second) Bose gas, $\Delta_1=2MT_1[1-e^{-2t_1/\tau_1}(1-\frac{\Delta}{2MT_1})]$ and the characteristic time of the $i$th Bose gas is $\tau_i=1/(\Gamma_i T_i^n)$ for $i=\{1, 2\}$.

At time $t=t_1+t_2$, the two temperatures can be simultaneously measured by the momentum distribution. The estimation precision of $(T_1,T_2)$ is governed by its covariance matrix $\textmd{Cov}(T_1,T_2)$, which is lower bounded via the multi-parameter Cram\'{e}r-Rao bound\cite{lab37}
\begin{align}
\textmd{Cov}(T_1,T_2)\geq\frac{1}{\mathbf{\chi}},
\end{align}
where $\mathcal{\chi}$ is the Fisher information matrix, which is derived by
 \begin{align}
\mathcal{\chi}_{jk}=\int_{-\infty}^\infty dP\frac{\partial_{T_j}f(t,P)\partial_{T_k}f(t,P)}{f(t,P)},
\end{align}
where $j, k=\{1, 2\}$.
The total simultaneous uncertainty of the two temperatures is achieved by
\begin{align}
(\delta^2T_1+\delta^2T_2)|_\textmd{sim}=\textmd{tr}[\textmd{Cov}(T_1,T_2)]\\
\geq\frac{\mathcal{\chi}_{11}+\mathcal{\chi}_{22}}{\mathcal{\chi}_{11}\mathcal{\chi}_{22}-|\mathcal{\chi}_{12}|^2}
\end{align}

The simultaneous uncertainty of the two temperatures can be analytically derived by the above equation. Especially, when $\Delta=0$, $P_0\neq0$, $T_1=T_2=T$ and $t_j=\tau_j$,  the simultaneous estimation precision of the two low temperatures is given by
\begin{align}
(\delta^2T_1+\delta^2T_2)|_\textmd{sim}
\geq 6.89625 T^2
\end{align}
This result shows that the simultaneous estimation of the two temperatures can also obtain very high measurement precision, which is proportional to  $T^2$ when both temperatures are close to $T$. However, the scaling $T^3$ can not be obtained like the measurement precision of a single temperature. This is mainly due to the fact that the variance of the momentum distribution, $\Delta_1/2$, after passing through the first Bose gas is not 0. Therefore, by controlling the initial momentum distribution, the measurement precision obtained by two separate temperature measurements is higher than that obtained by the simultaneous two-temperature measurement.

\section{conclusion}
In this article, we have proposed that the temperature of the Bose gas can be measured by the mobile heavy impurity. Due to the scattering off thermally excited quasiparticles in the Bose gas, the momentum distribution of the heavy impurity will carry the information about the temperature of the Bose gas.
Based on the Fisher information from the momentum distribution of the heavy impurity, the optimal measurement precision can be analytically derived.
Due to that the momentum is continuous, the Landau bound to precision $\delta^2 T\propto T^2$ can be obtained in the general case. And we find that enhancing the expectation value of the initial momentum can help to improve the measurement precision. Reducing the uncertainty of the initial momentum distribution can further improve the measurement precision of the temperature. More importantly, when the initial momentum
of the heavy impurity is fixed and non-zero, we obtain the measurement precision beyond the Landau bound to be $\delta^2 T\propto T^3$. In addition, we show that the momentum measurement is the optimal measurement when the initial momentum is fixed and not equal to 0. The kinetic energy measurement just happens to be the optimal measurement when the expectation value of the initial momentum is equal to 0. Finally, we find that the simultaneous measurement precision of the temperatures of the two Bose gases can also reach the Landau bound. However, the scaling $T^3$ can not be obtained like the measurement precision of a single temperature.

Our scheme can be performed in the setup with cold atoms such as, a one-dimensional quantum liquid of
$^4$He atoms confined within a porous material\cite{lab38}, and an ultracold Rb gas with single neutral Cs impurity atoms\cite{lab39}. The interaction strength between the impurity and host
atoms is tunable by the Feshbach resonances and powerful measurement techniques have been developed\cite{lab40,lab41}.

\section*{Appendix}
\textbf{A. The solution of the Fokker-Planck form}

The the Fokker-Planck form can be rewritten as
\begin{align}
\frac{\partial f(t,P)}{\partial t}=-\frac{1}{\tau}\mathcal{H}f(t,P),
\tag{A1}
\label{eq:A1}
\end{align}
where $\tau=|P/F|=\Gamma T^n$ and the operator $\mathcal{H}$ is described by
\begin{align}
\mathcal{H}=-MT\frac{\partial^2}{\partial P^2}-\frac{\partial}{\partial P}P.\tag{A2}
\end{align}
The general form of the solution  can de described as
\begin{align}
f(t,P)=e^{-\mathcal{H}t/\tau}f(0,P).\tag{A3}
\end{align}

Defining $\tilde{y}=P/\sqrt{2MT}$ and $\tilde{p}=-i\frac{\partial}{\partial P}$, the operator $\mathcal{H}$  is rewritten as
\begin{align}
\mathcal{H}=\frac{1}{2}(\tilde{p}-i\tilde{y} )^2+\frac{1}{2}\tilde{y}^2-\frac{1}{2}.\tag{A4}
\end{align}
Then, let us make a transformation to get an operator $\mathcal{\tilde{H}}$, which is similar with the Hamiltonian of a harmonic oscillator
\begin{align}
\mathcal{\tilde{H}}=\exp(\tilde{y}^2/2)\mathcal{H}\exp(-\tilde{y}^2/2)=\frac{1}{2}\tilde{p}^2+\frac{1}{2}\tilde{y}^2-\frac{1}{2}.\tag{A5}
\end{align}

The eigenvalues ($E_n$) and eigenvectors ($\phi_n(\tilde{y})$) of the operator $\mathcal{\tilde{H}}$ can be given by
\begin{align}
E_n=n, \phi_n(\tilde{y})=(-1)^n(\frac{1}{\sqrt{\pi}2^nn!})^{1/2}e^{y^2/2}\frac{d^n}{dy^n}e^{-y^2}.\tag{A6}
\end{align}
Setting $f(t,P)=e^{-\tilde{y}^2/2}\tilde{f}(t,P)$, we can obtain that
\begin{align}
\frac{\partial \tilde{f}(t,\tilde{y})}{\partial t}=-\frac{1}{\tau}\mathcal{\tilde{H}}\tilde{f}(t,\tilde{y}).\tag{A7}
\end{align}

The solution of above equation can be achieved by
\begin{align}
\tilde{f}(t,\tilde{y})=e^{-\mathcal{\tilde{H}}t/\tau}\tilde{f}(0,\tilde{y})\tag{A8}\\
=\sum_n e^{-\mathcal{\tilde{H}}t/\tau}A_n\phi(\tilde{y})\tag{A9}\\
=\sum_n e^{-n t/\tau}A_n\phi(\tilde{y}),\tag{A10}
\label{eq:A10}
\end{align}
where the coefficient $A_n$ is derived by
\begin{align}
A_n=\int_{-\infty}^\infty\tilde{f}(0,P)\phi^*(\tilde{y})dy.\tag{A11}
\end{align}

We consider that the initial momentum distribution is Gaussian, which is described by
\begin{align}
f(0,P)=\frac{1}{\sqrt{\pi\Delta}}\exp[-\frac{(P-P_0)^2}{\Delta}].\tag{A12}
\end{align}
The corresponding momentum distribution $f(0,\tilde{y})$ is given by
\begin{align}
f(0,\tilde{y})=\frac{1}{\sqrt{\pi\Delta'}}\exp[-\frac{(\tilde{y}-{P'_0})^2}{\Delta'}],\tag{A13}
\end{align}
where $\Delta'=\frac{\Delta}{2MT}$ and $P'_0=P_0/\sqrt{2MT}$. Next, the initial momentum distribution after the transformation is given by
 \begin{align}
\tilde{f}(0,\tilde{y})=\frac{1}{\sqrt{\pi\Delta'}}\exp[-\frac{(\tilde{y}-{P'_0})^2}{\Delta'}+\frac{\tilde{y}^2}{2}].\tag{A14}
\end{align}
Expanding it with the eigenvectors of the harmonic oscillator operator $\mathcal{\tilde{H}}$, the expansion coefficient can be obtained
 \begin{align}
A_n&=\int_{-\infty}^\infty\frac{1}{\sqrt{\pi\Delta'}}\exp[-\frac{(\tilde{y}-{P'_0})^2}{\Delta'}+\frac{\tilde{y}^2}{2}]\phi^*(\tilde{y})dy\tag{A15}\\
&=\int_{-\infty}^\infty d\omega\frac{(i\omega)^n\exp{[-\frac{\omega^2}{4}-\frac{\Delta'\omega^2+4iP'_0\omega-4{P'_0}^2}{4(1-\Delta')}]}}{\sqrt{4\pi\sqrt{\pi}2^nn!(1-\Delta')}},\tag{A16}
\end{align}
where we have utilized the following formulas
\begin{align}
e^{-{\tilde{y}^2}}&=\frac{1}{2\sqrt{\pi}}\int_{-\infty}^\infty d\omega e^{i\omega\tilde{ y}-\omega^2/4}\tag{A17}\\
\frac{d^n}{d\tilde{y}^n}(e^{-{\tilde{y}^2}})&=\frac{1}{2\sqrt{\pi}}\int_{-\infty}^\infty d\omega(i\omega)^ne^{i\omega\tilde{ y}-\omega^2/4}.\tag{A18}
\end{align}

Substituting the above equations into Eq.~(\ref{eq:A10}) and making an inverse transformation, we can obtain the solution of the Fokker-Planck formula in Eq.~(\ref{eq:A1})
\begin{align}
f(t,P)=&\frac{1}{\sqrt{2MT(1-e^{-2t/\tau}(1-\Delta'))}}\nonumber\\
&\exp[-\frac{(P-P_0e^{-t/\tau})^2}{2MT(1-e^{-2t/\tau}(1-\Delta'))}].\tag{A19}
\end{align}

\textbf{B. General form of Fisher information}

For the general initial Gaussian momentum distribution, the general form of the Fisher information can be analytically achieved by substituting Eq.~(\ref{eq:8}) into Eq.~(\ref{eq:10})
\begin{align}
\mathcal{F}(T)=&2\Gamma^2[T(2RMT+\Delta)]^{-2}\{n^2t^2T^{2n}[(P_0^2-4MT)\Delta\nonumber\\
&+2MT(RP_0^2+2MT)+\Delta^2]+(RMT/\Gamma)^2\nonumber\\
&+2nMRtT^{n+1}(2MT-\Delta)/\Gamma\},\tag{A20}
\end{align}
where $R=e^{2t\Gamma T^n}-1$.

\section*{Acknowledgements}
This research was supported by the National Natural Science Foundation of China (Grant No. 12365001 and No. 62001134), Guangxi Natural Science Foundation ( Grant No. 2020GXNSFAA159047), and the Project of Improving the Basic Scientific Research Ability of Young and Middle-aged Teachers in Universities of Guangxi (Grant No. 2023KY0815).

\end{document}